# NEW MANIFESTATIONS IN LOW-ENERGY ELECTRON SCATTERING FROM THE LARGE ACTINIDE ATOMS Cm and No

## Alfred Z. Msezane and Zineb Felfli

Department of Physics and CTSPS, Clark Atlanta University, Atlanta, Georgia 30314, USA

## Abstract

The Regge-pole calculated low-energy electron elastic total cross sections (TCSs) for Cm and No, characterized generally by negative-ion formation, shape resonances and Ramsauer-Townsend(R-T) minima, exhibit atomic and fullerene molecular behavior near threshold. Also, a polarization-induced metastable cross section with a deep R-T minimum near threshold is identified in the Cm TCSs, which flips over to a shape resonance appearing very close to threshold in the TCSs for No. We attribute these novel manifestations to size effects and orbital collapse impacting significantly the polarization interaction. This provides a new mechanism of tuning a shape resonance and R-T minimum through the polarization interaction. The comparison between the Regge-pole calculated ground, metastable and excited states anionic binding energies(BEs) with the existing theoretical electron affinities(EAs) demonstrates that those calculations tend to obtain metastable and/or excited states BEs and equate them incorrectly with the EAs, leading to ambiguous and unreliable electron affinity determination. A definitive and unambiguous meaning of the EA of complex heavy systems is recommended.

1. ## Introduction

In [1] the low-energy electron scattering from the actinide atoms Th, Pa, U, Np and Pu was investigated through the elastic total cross sections (TCSs) calculations. The objective was to delineate and identify the characteristic resonance structure in the TCSs as well as to understand and assess the reliability of the existing theoretical electron affinities (EAs). The understanding of chemical reactions involving negative ions necessitates the availability of accurate and reliable atomic and molecular affinities [2]. Indeed, the EA provides a stringent test of the theory when the calculated results are compared with those from reliable measurements. To our knowledge, there are no EA measurements available for the actinide atoms; their radioactive nature makes them difficult to handle experimentally. However, the recent experimental determination with high precision of the binding energy (BE) of the least-bound electron in atomic No [3] promises measurements of the EA as well.

Particularly interesting in the study [1] is the finding for the first time that the TCSs for atomic Pu exhibited fullerene molecular behavior [4,5] near threshold through the TCS of the highest excited state, while maintaining the atomic character through the ground state TCS. It is in this context that here we explore the low-energy electron scattering from the Cm and No atoms through the TCSs calculations to discover new manifestations. The data from the present and previous investigations of the actinide atoms and the already studied fullerene molecules $C_n(n=20,\ldots,112)$ [4,6] should contribute to a better understanding of the role of the individual atom/fullerene in the following. The study of the electronic structure and stabilization of $C_{60}$ through the encapsulation of the actinide atoms $An@C_{60}$(An=Th, ......., Md) [7]. Following its experimental preparation [8], the $Th@C_{76}$ has been characterized [9]. In the study of $An@C_{40}$ ( An = Th, ......., Md) [10], it was concluded that some of the clusters could be very stable; therefore, they could be useful in medicine and/or nuclear waste disposal. Also, the $M@C_{60}$ (M = Ti, Zr, U) fullerene hybrids have demonstrated catalytic efficiency in fundamental hydrogenation [11].

For most of the lanthanide atoms, producing sufficient anions that can be used in photodetachment experiments is very challenging [12]; for the actinide atoms the situation is even worse since their radioactive nature makes them difficult to handle experimentally. Theoretically, the large number of electrons involved and the presence of open d- and f-sub-shell electrons in both the lanthanide and actinide atoms, results in many intricate and diverse electron configurations. These lead to computational complexity of the electronic structure calculation, making it very difficult to obtain unambiguous and reliable EAs for these systems using structure-based theoretical methods. Indeed, many existing experimental measurements and sophisticated theoretical calculations have considered the anionic BEs of the stable metastable and/or excited negative ion formation to correspond to the EAs of the considered lanthanide and actinide atoms. This is contrary to the usual meaning of the EAs found in the standard measurement of the EAs of such complex systems as atomic Au, Pt and recently, At as well as of the fullerene molecules. In these systems, the EAs correspond to the ground state BEs of the formed negative ions.

Our robust Regge-pole methodology recently achieved a theoretical breakthrough through the identification of the electron-electron correlation effects and the core-polarization interaction as the major physical effects mostly responsible for negative ion formation in low-energy electron scattering from complex heavy systems. The novelty and generality of the Regge-pole approach is in the extraction of the anionic binding energies (BEs) from the calculated TCSs of the complex heavy systems; for ground state collisions these BEs yield the unambiguous and definitive theoretically challenging measured EAs. Very recently, the ground state anionic BEs extracted from our Regge-pole calculated electron elastic TCSs for the fullerene molecules $C_{20}$ through $C_{92}$ have been found to match excellently the measured EAs [4,6]. For the complex heavy systems Au and Pt the agreement between our Regge-pole calculated ground-state anionic BEs and the measured EAs of Au [13-15] and Pt [13, 16, 17] as well as of $C_{60}$ [18,19] is outstanding.

The proliferation in the literature of ambiguous and confusing meaning of the EAs of the lanthanide and actinide atoms requires that we first place this investigation in perspective. Recently, the EA of atomic Eu was measured to be $0.116 \pm 0.013$ eV [12], which is in outstanding agreement with the Regge-pole [20] and MCDF-RCI [21] calculated values. Also, the previously measured EA value of $1.053 \pm 0.025$ eV for Eu [22] agrees excellently with the Regge-pole value of 1.08 eV [23]. However, the Regge-pole values correspond to the BEs of metastable/excited Eu⁻ negative ions [24]. The conundrum in the measured EAs of Eu has been discussed and resolved recently [24]. For the Tm atom the measured EA value of 1.029 eV [25] also agrees excellently with our anionic metastable state BE of 1.02 eV [24]. These results give great credence to the power and ability of the Regge-pole methodology to produce reliable ground, metastable and excited state BEs of complex heavy systems through the TCSs calculation. Indeed, the Regge-pole methodology requires no assistance whatsoever from either experiment or other theory to achieve the remarkable feat.

## 2. Method of calculation

### 2.1 Elastic scattering total cross section (TCS)
Most existing theoretical methods used for calculating the anionic BEs of complex heavy systems are structure-based. Regge poles, singularities of the S-matrix, rigorously define resonances [26,27]; consequently, they are appropriate for use in this investigation. Being generalized bound states within the complex angular momentum (CAM) description of scattering, the Regge-poles can be used to determine reliably the anionic BEs of the ground, metastable and excited states of complex heavy systems, provided the essential physics has been accounted for adequately as in our case here. In the Regge-pole, also known as the CAM methods the important and revealing energy-dependent Regge trajectories are also calculated. Their effective use in low-energy electron scattering has been demonstrated in for example [20,28].

The Mulholland formula [29] is used here to calculate the near-threshold electron–atom/fullerene collision TCS resulting in negative ion formation as resonances. In the form below, the TCS fully embeds the essential electron-electron correlation effects [30,31] (atomic units are used throughout):

$$\sigma_{tot}(E) = 4\pi k^{-2} \int_0^\infty \text{Re}[1 - S(\lambda)]\lambda d\lambda$$
$$- 8\pi^2 k^{-2} \sum_n \text{Im} \frac{\lambda_n \rho_n}{1 + \exp(-2\pi i \lambda_n)} + I(E) \quad (1)$$

In Eq. (1) $S(\lambda)$ is the S-matrix, $k = \sqrt{2mE}$, $m$ being the mass and $E$ the impact energy, $\rho_n$ is the residue of the S-matrix at the $n^{th}$ pole, $\lambda_n$ and $I(E)$ contains the contributions from the integrals along the imaginary $\lambda$-axis ($\lambda$ is the complex angular momentum); its contribution has been demonstrated to be negligible [20].

### 2.2 The Potential
As in [32] here we consider the incident electron to interact with the complex heavy system without consideration of the complicated details of the electronic structure of the system itself. Therefore, within the Thomas-Fermi theory, Felfli *et al* [33] generated the robust Avdonina-Belov-Felfli (ABF) potential which embeds the vital core-polarization interaction

$$U(r) = -\frac{Z}{r(1+\alpha Z^{1/3} r)(1+\beta Z^{2/3} r^2)} \quad (2)$$

In Eq. (2) $Z$ is the nuclear charge, $\alpha$ and $\beta$ are variation parameters. Note also that the ABF potential has the appropriate asymptotic behavior, *viz.* $\sim -1/(\alpha\beta r^4)$ and accounts properly for the polarization interaction at low energies. The strength of this extensively studied potential [34-36] lies in that it has five turning points and four poles connected by four cuts in the complex plane. The presence of the powers of Z as coefficients of $r$ and $r^2$ in Eq. (2) ensures that spherical and non-spherical atoms and fullerenes are correctly treated. Also appropriately treated are small and large systems. The effective potential $V(r) = U(r) + \lambda(\lambda+1)/2r^2$ is considered here as a continuous function of the variables $r$ and complex $\lambda$. The details of the numerical evaluations of the TCSs have been described in [31] and further details of the calculations may be found in [37].

The novelty and generality of the robust Regge-pole approach lies in that the crucial electron-electron correlation effects are fully embedded in the Mulholland formula. And the ABF potential contains the vital core-polarization interaction. These two important effects have been identified as the major physical effects mostly responsible for electron attachment in low-energy electron scattering from complex heavy systems, leading to stable negative ion formation. Consideration of these effects is particularly important in the study of the experimentally difficult to handle radioactive actinide atoms and because of the inability of conventional theoretical calculations to obtain unambiguous EAs for the lanthanide atoms [24]. Also crucial in the CAM methods are the Regge trajectories, viz. Im $\lambda_n(E)$ versus Re $\lambda_n(E)$ ($\lambda_n(E)$ is the CAM). These probe electron attachment at the fundamental level near threshold, thereby allow for the determination of reliable EAs. Thylwe [28] investigated Regge trajectories, using the ABF potential, and demonstrated that for Xe atom the Dirac Relativistic and non-Relativistic Regge trajectories yielded essentially the same Re $\lambda_n(E)$ when the Im $\lambda_n(E)$ was still very small. This clearly demonstrates the insignificant difference between the Relativistic and non-Relativistic calculations at near threshold electron impact energies, provided the appropriate physics is accounted for adequately as in our case.

The potential (2) has been used successfully with the appropriate values of $\alpha$ and $\beta$. It has been found that when the TCS as a function of $\beta$ has a resonance [20], corresponding to the formation of a stable negative ion, this resonance is longest lived for a given value of the energy, which corresponds to the EA of the system (for ground state collisions) or the BE of the metastable/excited anion. This was found to be the case for all the systems, including fullerenes we have investigated thus far. This fixes the optimal value of "$\beta$" in Eq. (2) when the optimum value of $\alpha = 0.2$. The effective use of Im $\lambda \to 0$ is demonstrated in [20] and carefully explained in [38]. Although we have previously referred to Connor [39] for the physical interpretation of Im $\lambda(E)$, the original interpretation was given by Regge himself [40]. For a true bound state, namely $E < 0$, Im $\lambda(E) = 0$ and therefore the angular life, $1/(Im \lambda(E)) \to \infty$, implying that the system can never decay. Obviously, in our calculations Im $\lambda(E)$ is not identically zero, but small – this can be clearly seen in the figures [38]: the dramatically sharp (long-lived) resonances hardly have a width as opposed to shape resonances for instance (see also [20] for comparison). We limit the calculations of the TCSs to the near threshold energy region, namely below any excitation thresholds to avoid their effects.

We emphasize here that the investigation of the Regge trajectories allows us to readily see the electron impact energy range where Relativity is unimportant in the calculation of the TCSs, i.e. at small electron impact energies. Also, the use of the Im $\lambda_n(E)$ allows us to differentiate among the important ground and metastable negative ion formation as well as the shape resonances.

The very small value of Im $\lambda_n(E)$ indicates the ground state negative ion formation.

## 3. Results

Figures 1 and 2 present the electron collision TCSs for Cm and No atoms, respectively. Indeed, these TCSs are characterized by ground, metastable and excited states negative ion formation. At a glance the TCSs for each atom in the Figures appear complicated. However, they are readily understood and interpreted if we focus on a single color-coded curve at a time since it represents scattering from different states resulting in negative ion formation. Generally all the TCSs in the Figures are characterized by R-T minima, shape resonances (SRs) and dramatically sharp resonances corresponding to the Cm⁻ and No⁻ anionic formation. Consequently, the need for their delineation and identification is evident so that the unambiguous determination of the EAs of both Cm and No can be obtained.

The fundamental physics underlying these curves can be understood readily if we focus on each color-coded TCS. For the analysis we select the ground state TCS curve, the red curve of Fig. 1. From near threshold the ground state TCS decreases monotonically with the increase of the electron impact energy, E reaching the first R-T minimum at about 1.1 eV. This behavior is typical of the standard TCSs for complex heavy atoms such as Au and Pt as well as others. With the further increase in energy, the electron becomes trapped by the centrifugal barrier manifested by the appearance of the shape resonance (SR) at about 1.6 eV. As the electron leaks out of the barrier, it polarizes the Cm atom strongly reaching maximum polarization at about 3.30 eV. Here the Cm atom is transparent to the incident electron leading to electron attachment forming the stable Cm⁻ anion with the BE value of 3.32 eV. The electron spends many angular rotations as the Cm⁻ anion decays, with the angular lifetime determined by $1/(\text{Im}\lambda(E)) \to \infty$, since for ground state anionic formation $\text{Im}\lambda(E) \to 0$.

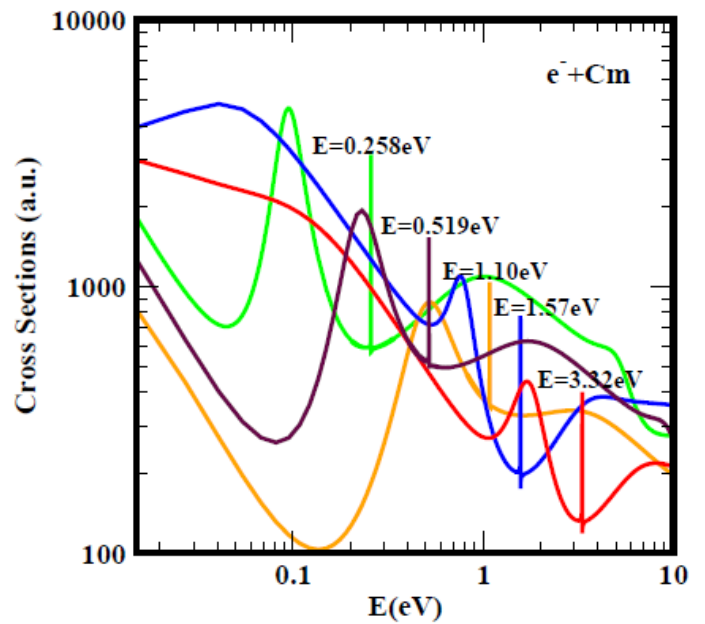

**Fig. 1:** Total cross sections (a.u) for electron-Cm scattering. Red, blue and orange curves represent respectively ground and metastable TCSs. Brown and green curves are two excited states TCSs. The orange curve is the polarization-induced TCS due to size effect.

Indeed, the results demonstrate the crucial importance of the polarization interaction in low-energy electron scattering calculations of TCSs of complex heavy systems. These characteristic R-T minima, also observed in the Dirac R-matrix low-energy electron elastic scattering cross sections calculations for the heavy alkali-metal atoms Rb, Cs and Fr [41], manifest that the important polarization interaction has been accounted for adequately in our calculation consistent with the conclusion by Johnson and Guet [42]. The energy positions of the sharp resonances correspond to the anionic BEs of the formed negative ions during the electron collision with the ground, metastable and excited Cm and No atoms. The data from the Figures 1 and 2 are summarized in Table 1. The anionic BEs for the ground, metastable and excited Cm⁻ and No⁻ negative ions are compared with the Relativistic Configuration-Interaction (RCI) [43] and the QR-LSD-GX-SIC-GWB [44] calculated EAs. It is noted that the calculated EA value for Cm using the RCI theory 0.321 eV [43] and using the VWN energy-correlation correction method 0.283 eV [44] compare reasonably well with the Regge-pole calculated anionic BE of the highest excited state, EX-1. However, the calculated EA value of 0.449 eV using the SPP energy-correlation correction theory [44] is closer to the Regge-pole calculated BE of the second excited anionic state, EX-2.

For the No atom, the calculated absolute EAs using the SPP and the VWN energy-correlation correction theories [44] are very close to each other and comparable to the Regge-pole calculated BE value of 1.92 eV. However, this value corresponds to the anionic BE of the first metastable state and not the ground state BE which is large, viz. 3.83eV. It is noted here that the RCI calculation could not obtain a converged EA value for the No atom. Clearly, these theoretical results provide the strongest evidence of the ambiguous and uncertain EA determination of the Cm and No actinide atoms. As already pointed out in the Introduction, these EA values reflect the inability of the sophisticated theoretical methods to obtain unambiguous and definitive EA values of the actinide atoms as well as the lanthanide atoms, see also Table 1 of [5].

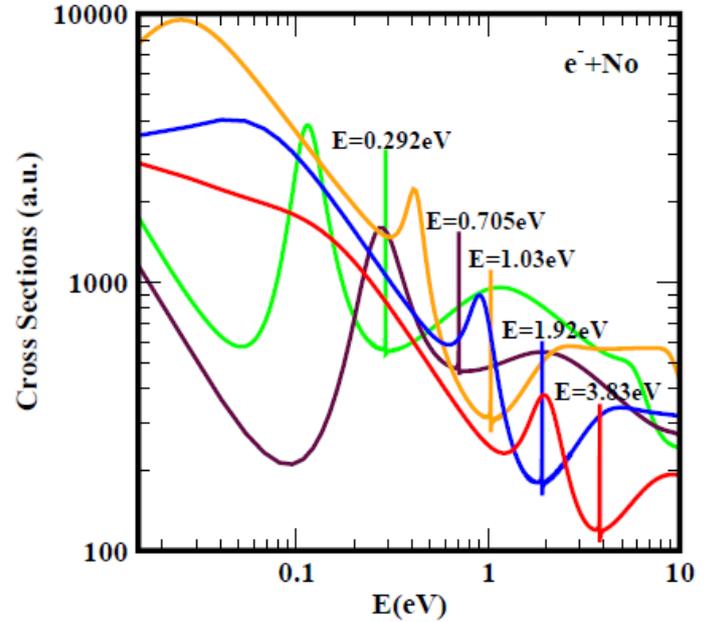

**Fig. 2:** Total cross sections (a.u) for electron-No scattering, red, blue and orange curves represent respectively ground and metastable TCSs. Brown and green curves are two excited states TCSs. The orange curve is the polarization-induced TCS due to size effect.

**Table 1:** Negative ion binding energies (BEs), in eV obtained from the TCSs for the actinide atoms Cm and No. GRS, MS-*n* and EX-*n* (*n*=1, 2) represent respectively ground, metastable and excited states. The experimental EAs, EXPT and theoretical EAs, Theory are also indicated; N/A denotes not available. Table 1 demonstrates the importance of determining the ground state BEs of the formed anions during the collisions. R-T Min is the energy position of the ground state R-T minimum in eV.

| Z | Atom | BEs GRS | EAs EXPT | BEs MS-1 | BEs MS-2 | BEs EX-2 | BEs EX-1 | EAs Theory | R-T Min GRS |
|---|------|---------|----------|----------|----------|----------|----------|------------|-------------|
| 96 | Cm | 3.32 | N/A | 1.57 | 1.10 | 0.519 | 0.258 | 0.321[43] 0.283[44] 0.449[44] | 3.30 |
| 102 | No | 3.83 | N/A | 1.92 | 1.03 | 0.705 | 0.292 | -2.301[44] -2.325[44] | 3.85 |

We now explain the atomic/fullerene molecular behavior in the TCSs of Cm and No and the transformation of the polarization-induced R-T minimum in the Cm TCSs to a shape resonance near threshold in those for No. In [4] atomic/fullerene molecular behavior was investigated through the calculation of the TCSs for $C_{20}$ through $C_{112}$. For this we focus on the highest excited states curves, the green curves in both the Cm and No TCSs. In [4] the ratio of the second to the first R-T minima determined atomic/fullerene molecular behavior. For atomic behavior this ratio was greater than unity; a ratio less than unity corresponded to fullerene molecular behavior. Indeed, although the TCSs of Fig. 1 are for the Cm atom, this ratio is less than unity, indicative of fullerene molecular behavior, see also the curves of Ref. [4]. Another important manifestation in the TCSs of Fig. 1 is the appearance of the important polarization-induced metastable TCS with the deep R-T minimum at 0.12 eV.

The TCSs of Fig. 2 behave significantly differently from those of Fig. 1. The highest excited state TCS of Fig. 2 has a ratio of about unity, indicative of neither atomic nor fullerene molecular behavior (its behavior resembles that of the $C_{24}$ fullerene molecule [4]). Secondly, the polarization-induced metastable TCS of Fig. 1 (orange curve) has flipped over from R-T minimum at 0.12eV to a shape resonance at about 0.024eV close to threshold. This behavior and the large ground state anionic BE of 3.83eV reflect the importance of the high Z of No compared to that of Cm. Indeed, the rich in resonance structures, including R-T minima of the TCSs of both Cm and No require the careful delineation and identification of the various dramatically sharp resonances in these TCSs for unambiguous and definitive determination of the EAs. Clearly, the incorrect identification of the sharp peaks in the TCSs has led to the proliferation of unreliable EAs for the Cm and No atoms and, generally of the actinide atoms.

It is worth remarking on the impact of the 6d-orbital collapse on the anionic BEs when transitioning from $Cm(7s^25f^76d^1)$ to $Bk(7s^25f^9)$. It impacts the core-polarization interaction by increasing the ground and first excited anionic states BEs of Cm⁻ from 3.32eV and 1.57eV to respectively 3.55eV and 1.73eV in the Bk⁻ anion. Notably, the ground and the lowest metastable BEs of the Cm⁻ anion are essentially the same as those of the Pu⁻ and Am⁻ anions, see also Table 1 of [1]. Incidentally, the impact of the 5d-orbital collapse on the polarization interaction has already been discussed in the context of the negative ion BEs of the lanthanide atoms[45].

## 4. Summary and Conclusions

Here we have probed the response of the large actinide atoms Cm and No to low-energy electron collisions through the elastic TCSs calculations using our robust Regge-pole method. Embedded in the method are the crucial electron correlation effects and the vital core-polarization interaction; these are the major physical effects responsible for electron attachment in complex heavy systems, leading to stable negative ion formation. The objective was to discover new manifestations and understand the existing theoretical EAs for these atoms as well as present their first time TCSs. The TCSs for both Cm and No atoms were found to be characterized generally by ground, metastable and excited negative ion formation, shape resonances and R-T minima. The extracted from the TCSs anionic BEs were compared with the existing theoretical EAs; the experimental EAs are unavailable because of the difficulty of handling the actinide atoms experimentally due to their radioactive nature.

In conclusion, we have discovered new manifestations in the low-energy electron scattering TCSs of both Cm and No atoms; namely, atomic and fullerene molecular behavior near threshold. Also, a polarization-induced metastable cross section with a deep R-T minimum near threshold has been identified in the Cm TCSs, which flips over to a shape resonance appearing very close to threshold in the TCSs for No. We have attributed these peculiar tunable behaviors in the TCSs to size effects impacting significantly the polarization interaction. This provides a novel mechanism of tuning a shape resonance and R-T minimum through the polarization interaction via the size effect. The comparison between the Regge-pole calculated ground, metastable and excited states anionic BEs with the existing theoretical EAs demonstrates that the existing theoretical calculations tend to obtain metastable and/or excited states BEs and equate them incorrectly with the EAs. Indeed, this leads to ambiguous and unreliable determination of the EAs of complex heavy systems; they are already populating the literature. For a definitive and unambiguous meaning of the EA, we recommend the use of the ground state anionic BE as the EA of complex heavy systems, consistent with the use in the determination of the EAs of such atoms as the complex heavy Au, Pt and At as well as of the fullerene molecules.

Finally, the obtained ground, metastable and excited negative ion BEs for the Cm and No atoms can now be used in sophisticated theoretical methods such as the Dirac R-matrix, MCHF, Relativistic Configuration Interaction, etc. to construct the appropriate wave functions and for fine-structure energy calculations. Particularly important here for reliable calculations and measurements as well is the determination of the ground states anionic BEs of the formed negative ions during the collisions. Perhaps, future measurements of the EAs of complex heavy systems could be carried out using the electron transmission method [46] for their definitive determination.

**Acknowledgments**
Research was supported by the U.S. DOE, Division of Chemical Sciences, Geosciences and Biosciences, Office of Basic Energy Sciences, Office of Energy Research. The computing facilities of the National Energy Research Scientific Computing Center, also funded by U.S. DOE are greatly appreciated.